\documentclass{aa}

\usepackage{graphicx}
\usepackage[varg]{txfonts}
\usepackage{amsmath}
\begin{document}

   \title{A Wide-band High-frequency Type-II Solar Radio Burst}


   \author{V. Vasanth \inst{1}
          \and
          Yao Chen \inst{2}
          \and
          G. Michalek \inst{1}
          }

   \institute{Astronomical Observatory of Jagiellonian University,
              Orla 171, Krakow 30-244, Poland\\
              \email{vasanth.veluchamy@uj.edu.pl}
         \and Institute of Space Sciences, Shandong University, Shandong, 264209, China.
             \email{yaochen@sdu.edu.cn}
             }

   \date{Received ; accepted}

  \abstract

   {Type-II radio bursts are typically observed below $\sim$ 400 MHz, with narrow-band slowly-drifting fundamental and harmonic structures. Here we report an unusual high-frequency wide-band type-II burst with starting frequency as high as 600$-$700 MHz and the instantaneous bandwidth being as wide as $\sim$ 300 MHz.}
   {We used radio imaging from Nan{\c{c}}ay Radio Heliograph, spectroscopic data from ORFEES, and extreme-ultraviolet observation from Solar Dynamics Observatory and white-light observations from LASCO to determine the nature and origin of the observed radio burst as well as their propagation in the corona.}
   {The estimated average spectral drift is $\sim$ 2.18 MHz s$^{-1}$, its mean duration at each frequency is $\sim$ 3 min, the maximum brightness temperature can exceed 10$^{11}$ to 10$^{12}$ K. According to the simultaneous EUV and radio imaging data, the radio sources distribute over a relatively broad region that concentrates around a dip of the nose front of the shock-like EUV wave structure. The dip is likely caused by the strong interaction of the eruption with the overlying closed dense loops that are enclosed by the large-scale streamer structure, indicating that the Type-II burst originates from such CME-shock interaction with dense-closed loop structures.}
   {The observations suggest that the unusual wide-band high frequency type-II radio burst originates from a dense streamer region in the corona, supported by EUV shock-like structure that get steepened very close to the solar surface $\sim$ 1.23 R$_\odot$ and type-II radio source coincides with the shock dip. Further the wide-band feature is due to the source stemming from a region with large density variation and not due to the intensity variation across the shock structure.}

   \maketitle
%

\section{Introduction}
  Type$-$II solar radio bursts were first identified by
Payne-scott, Yabsley and Bolton (1947). Such bursts are
characterized by narrow bands drifting slowly from higher
to lower frequencies according to the radio dynamic spectral
data. They are associated with energetic electrons
accelerated at coronal shocks (Wild, 1950; Wild and McCready, 1950;
Weiss, 1963; Smerd, 1970; Dulk, 1985; Nelson and Melrose, 1985 for earlier reviews).
Some bursts exhibit a pair of bands that represent the
fundamental (F) and harmonic (H) radiations according to the
classical theory of plasma emission (Ginzburg \& Zhelezniakov, 1958).
In some events the F and H bands may further split into two
well-separated yet morphologically-similar sub-bands,
being referred to as the band-split type-II burst (McLean 1967;
Wild \& Smerd 1972; Smerd et al. 1974). The exact emission
process of type-II burst and their splitting bands
are still under investigation (see, e.g., Vasanth et al., 2014; Du et al., 2014, 2015; Ni et al., 2020;
Chen et al. 2022a, 2022b).

The origin of Type-II bursts and their association
with eruptive structure/activities are a long-standing question
(see, e.g., Reiner et al. 2003; Mancuso and Abbo 2004; Cho et al. 2007;
Chen et al. 2013, 2014; Vasanth and Umapathy, 2013; Shanmugaraju et al. 2018;
Morosan et al. 2022; Vasanth, 2024 and many others).
According to these earlier studies both the shock flank and the shock nose
front can act as the sources of type-II bursts. One important
scenario suggests that the coronal type-II bursts
(at the metric-decimetric wavelength)
originate from the interaction of coronal mass ejection (CME)
with nearby dense structures such as coronal rays or streamers
(Reiner et al. 2003; Mancuso and Abbo 2004; Cho et al. 2007, 2011;
Chen et al. 2013, 2014; Kong et al. 2015, 2016; Chrysaphi et al. 2018, 2020).
Such interaction favors the generation of type-II radio bursts
as it may lead to: (1) local steepening or enhancement
of the shock structure since the rays or streamers are denser
than surroundings and thus with lower Alfv{\'e}nic speed,
(2) quasi-perpendicular shock geometry if the CME interacts
with overlying loops or the CME impinges
on the streamer/ray structure from its flank, (3) the transit of shock
across magnetically-closed structure that forms
an efficient particle trap ahead of the shock structure
leading to repetitive acceleration. Kong et al. (2012)
and Chen et al. (2014) presented the observational
evidence of CME-ray/streamer interaction region acting as
the source of type-II burst, and in a series of numerical studies
Kong et al. (2015, 2016) have verified the scenario that
the shock interaction with large-scale magnetically-closed
structure acts as efficient particle accelerator and potential type-II sources.

\begin{figure*}[ht]
\includegraphics[width=0.95\textwidth]{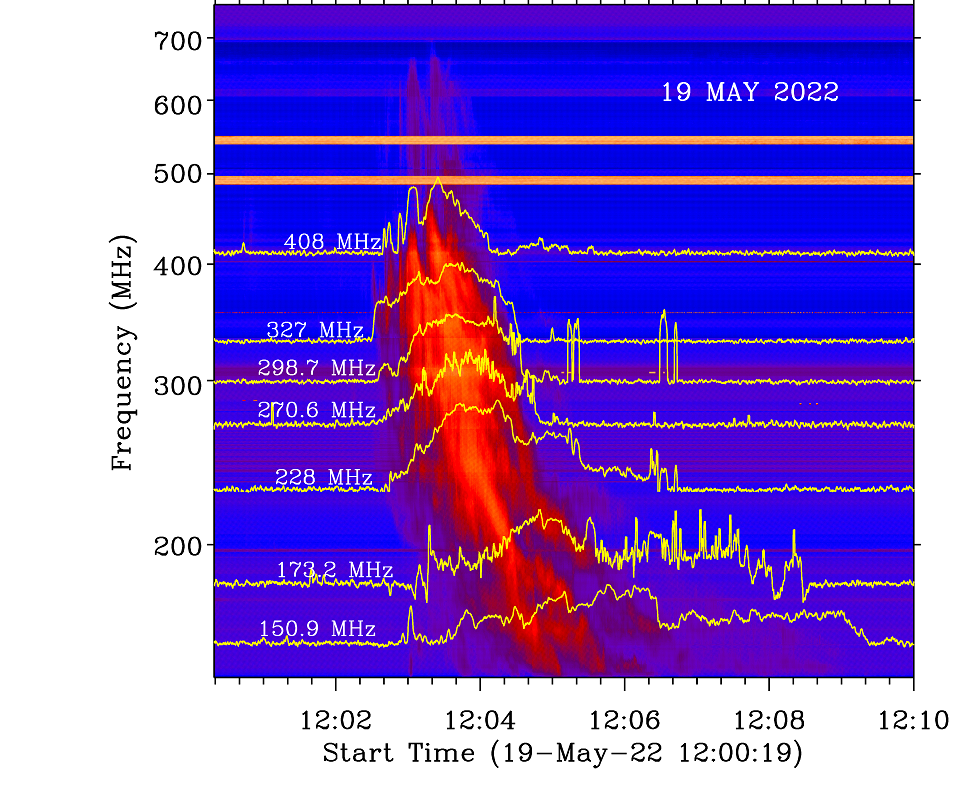}
\caption{The dynamic spectrum of the high frequency wide-band type-II burst recorded by ORFEES (600 $-$ 144 MHz) radio spectrograph.
The maximum brightness temperature ($T_{Bmax}$) at different NRH frequencies are plotted with yellow lines over the dynamic spectrum.}
\label{fig:fig1}
\end{figure*}

Reiner et al. 2003, analyzed a type-II radio burst accompanied by two CMEs
and suggested that the type-II burst is generated by CME shock propagating through streamer.
They don$^{\prime}$t have radio imaging observations, so they don$^{\prime}$t know the precise location
of radio source along the shock. Chrysaphi et al. 2020 analyzed a stationary
to drifting structured type-II radio burst accompanied by two CMEs and concluded that
streamer-puff CME is the driver of the type-II burst, further the shock interaction with
the streamer results in stationary component of the type-II burst and later expansion of
streamer is related to drifting component of the type-II burst. They have radio imaging
from LOFAR observations between 45$-$30 MHz corresponding to heliospheric distance of
$\sim$ $>$ 1.8 R$_\odot$, they showed their centroid of radio sources on the occulting
disk of LASCO CME and compared it with the eruptive structure in LASCO C2 FOV. The radio
sources are located at the CME flank. Similarly Chrysaphi et al. 2018 analyzed a
band-splitting radio burst from LOFAR observations with radio imaging in the same frequency
range between 45$-$30 MHz corresponding to heliospheric distance of $\sim$ $>$ 1.8 R$_\odot$.

Thus the previous studies (Reiner et al. 2003, Chrysaphi et al. 2018, 2020) lacks
radio imaging observations near the sun. The simultaneous observations of EUV
and radio imaging are available for our event. We take the advantage of NRH imaging observations
closer to the solar surface and provide the precise location of radio sources over the shock structure
at lower corona as well their streamer region for the first time for an unusual wide-band high
frequency type-II radio burst. It is to be noted that Feng et al. 2016 has analyzed the complex
coronal eruption and their radio bursts, reported a shift of radio sources from flank to front of the shock during
propagation likely due to change of shock geometry at different sides of the shock during its
evolution.

\begin{figure*}[ht]
\includegraphics[width=0.95\textwidth]{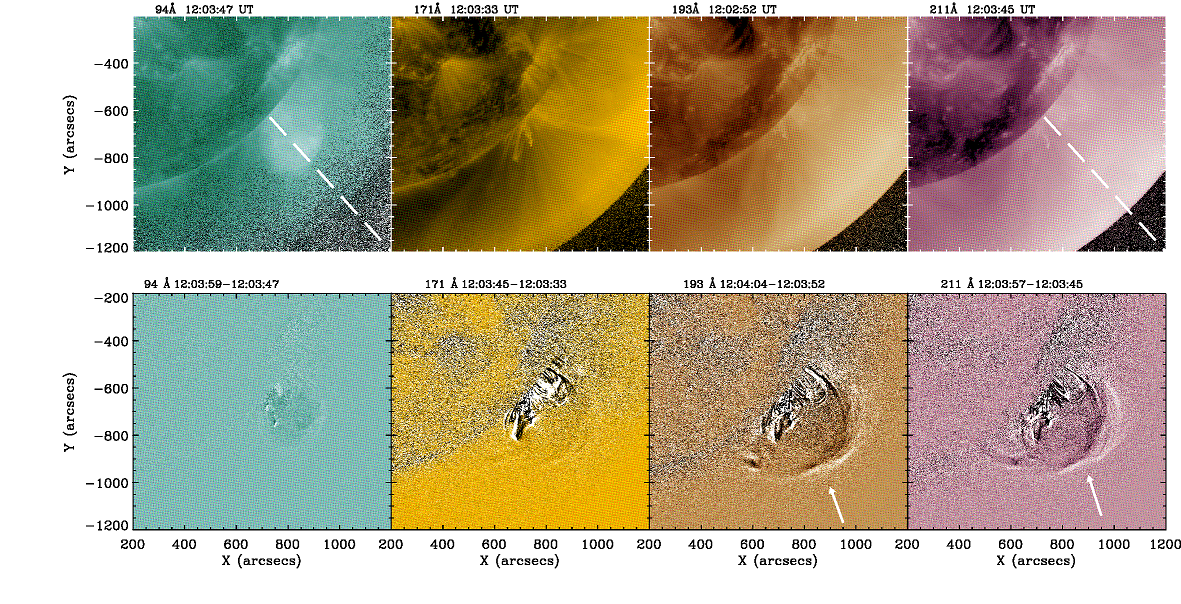}
\caption{The eruptive structures observed by AIA at 94, 131, 171, and 211 {\AA} (top panel), and the corresponding running difference images (bottom panel).
The dashed white line is the slit for distance map to be shown in Figure 4. The duration of the animation is 3s. (An animation of this figure is available).}
\label{fig:fig2}
\end{figure*}

Type-II events with the starting frequency higher
than $\sim$ 400 MHz are rare and have been reported earlier
(e.g., Pohjolainen et al., 2008; Cho et al. 2013; Kumari et al. 2017).
Such events may stem from the above-mentioned CME interaction
with dense coronal structures such as streamers/rays or dense loops,
or from sources that are relatively close to the solar surface.

Type$-$II bursts usually present narrow bands, with
a relative bandwidth ($\delta f $/$f$) being less than
0.1 - 0.5 (e.g, Mann et al. 1996; Vasanth et al. 2014). In some Type-II events
the relative band width can be larger than 0.5 (e.g., Feng et al. 2015;
Zimovets and Zadykov, 2015; Lv et al. 2017; Morosan et al. 2022, Zhang et al. 2023).
Two scenarios exist to explain the wide-band nature,
(1) the bursts originate from broad sources or separated locations along the front,
or (2) the sources are from the transition layer across the shock
with large density fluctuation.

Here we report an unusual event with both high starting frequency
and wide band. Both EUV and radio imaging data are available,
allowing us to further explore its origin. The
following section presents the spectral and imaging data of the event.
Section 3 shows the EUV and white light (WL) data of the eruption,
and Section 4 presents the combined analysis of these data.
The final section summarizes the study with discussion.

\begin{figure*}[ht]
\includegraphics[width=0.95\linewidth]{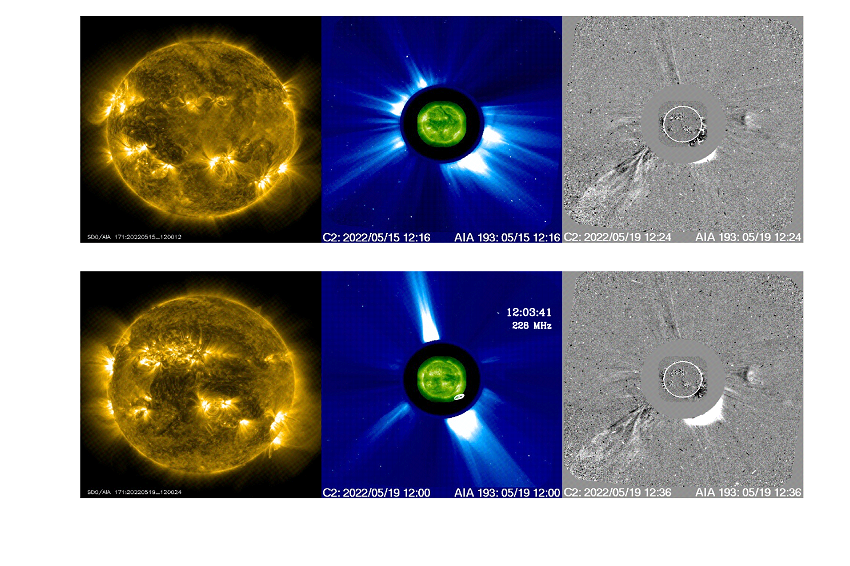}
\caption{The source active region observed on the solar limb on 15/05/2022 in AIA 171  {\AA}, and LASCO C2 images with possible CME-streamer interaction regions (top panel). The eruptive structure on the day of the type-II with LASCO images (bottom panel).The radio
sources are represented by the 50 $\%$, 70 $\%$, and 90 $\%$
contours of the corresponding T$_{Bmax}$.}
\label{fig:fig3}
\end{figure*}

\section{The Radio Imaging Spectroscopy Observation}

Figure 1 presents the spectral data of the type-II burst
recorded on 19 May 2022 by the ORFEES radio spectrograph,
from 144 to 1000 MHz with a time resolution of 0.1s. The
burst started at $\sim$ 12:02 UT with a duration of $\sim$ 22 mins.
The starting frequency is $\sim$ 600 MHz, yet later
some weaker emission appears at frequency up to 700 MHz.

The burst drifts downwards to $\sim$ 150 MHz within $\sim$ 2.5 mins,
with an average drift rate of $\sim$ 2.18 MHz s$^{-1}$,
measured from 11 frequency-time points.
Such value agrees with earlier studies on high
frequency type$-$II events (e.g., Pohjolainen et al. 2008;
Cho et al. 2013; Kumari et al. 2017), and being larger than those
reported for typical type-II bursts starting at lower frequency
($\le$ 1 MHz s$^{-1}$, see, e.g., Mann et al. 1996; Vrsnak et al. 2001;
Vasanth et al. 2011, 2014; Feng et al. 2013).

The present events consists of irregular band-split signature at
the start(before 12:03:20 UT) and later (after 12:05 UT).
During the main part of the burst no such signature can be identified.
Instead, the main part exhibits an overall enhancement across the
spectral band as wide as 300 MHz that is at the same order of the
emission frequency. The maximum intensity appears around the
middle part of the spectra, indicating that the wide-band part
is not due to the overlappings of splitting bands.
In addition, at any specific frequency, the type-II burst lasts
for $\sim$ 2 mins, also being much longer than usual bursts.
These observations are very interesting since normal type-II bursts
are characterized by slowly-drifting narrow bands. According
to the spectral data, the relative bandwidth
($\delta f $/$f$) is $\sim$ 1 $\pm$ 0.05 measured at
10 moments between 12:03-12:05 UT. For instance,
at 12:03:41 UT, the emission extends from $\sim$ 440 MHz
to $\sim$ 180 MHz, and the central frequency is $\sim$ 300 MHz, yields $\delta f $/$f$ = 1.

The type-II sources were imaged by Nan{\c{c}}ay Radio Heliograph
(NRH: Kerdraon and Delouis, 1997) at
several frequencies (408, 327, 298.7, 270.6, 228, 173.2, and 150.9 MHz).
Its spatial resolution depends on the imaging
frequency and time of the year, which is $\sim$ 1'-2' at
445 MHz and $\sim$ 5'-8' at 150.9 MHz for the present event.
Its time resolution is 0.25s and we used the integrated data of 1s for
a higher signal-noise ratio.

The temporal profiles of the brightness
temperature ($T_B$) obtained by NRH. The $T_B$ reaches up to $\sim$ 10$^{12}$ K
at frequencies above 228 MHz, at lower frequencies, $T_B$ reaches above 10$^{10}$ K.
 This agrees with the usual interpretation
of type-II burst in terms of coherent plasma emission.
\hfill \break

\section{The EUV and WL Data of the Eruption}

The type-II burst was associated with a CME originating
from the limb according to Atmospheric Imaging Assembly (AIA: Lemen et al. 2012)
onboard the Solar Dynamics Observatory (SDO: Pesnell et al. 2012)
spacecraft and Large Angle Spectrometric Coronagraph
(LASCO: Brueckner et al., 1995) C2 onboard the Solar and Heliospheric Observatory
(SOHO: Domingo, Fleck, and Poland, 1995). See Figure 2 and the accompanying
movie for the AIA data at 171, 193, 211, and 94 {\AA}.
The CME originated from the backside of the sun, the source
active region (NOAA AR 13006) and overlying loops were
well observed several days ago (see Figure 3).

The early stage of the CME was well observed by AIA.
 The ejecta is associated with
a highly-writhed multi-thermal structure, as seen
from the hot 94 {\AA} passband, the warm passbands
at 211, 193, and 171 {\AA}, as well as the cool
304 {\AA} passband (not shown here). The EUV wave
front appears ahead of the ejecta from the
running difference images at 193 and 211 {\AA}
(see Figure 2 and the accompanying movie).
The wave structure gets steepened after 12:02 UT. To
evaluate the speeds of the EUV wave and the underlying
ejecta we plot the height-time maps along the given slit
with the 211 and 94 {\AA} data (Figure 4).
The fitted average speed of the 94 {\AA} ejecta is $\sim$
703 $\pm$ 70 km s$^{-1}$, and that of the EUV wave is $\sim$ 868 $\pm$ 120 km s$^{-1}$
with the 211 {\AA} data.

The eruption appears in the LASCO C2 field of view
at 12:24 UT. There exist bright and dense loops underlying a bright streamer
structure, according to Figure 3 of the AIA and LASCO C2 data
right before the eruption and four days ago. So the CME originates
from the magnetically-closed loop system that is enclosed by
the large-scale streamer. Strong CME-shock interaction with
these overlying loops and the larger streamer structure
must occur later.

\begin{figure}
\hfill\break
\hfill\break
\hfill\break
\includegraphics[width=\hsize]{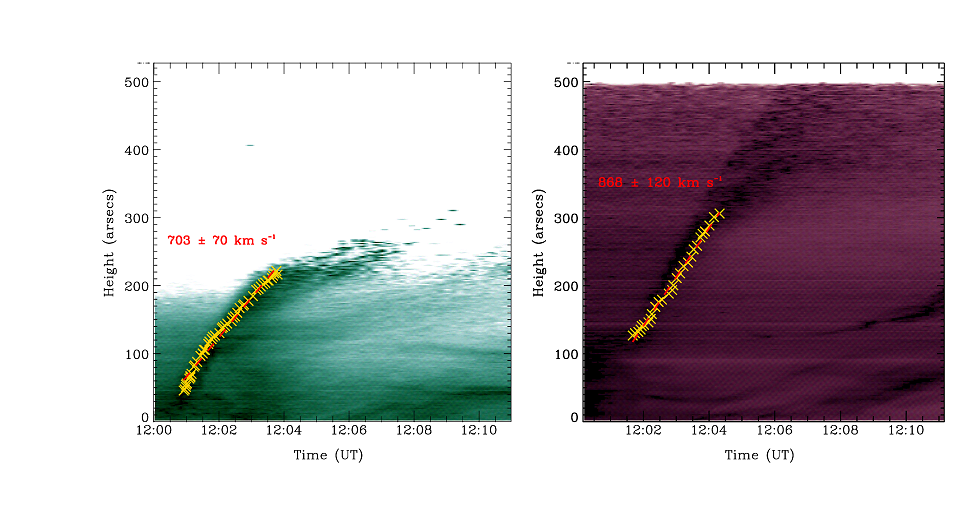}
\caption{Distance-time maps along the the slice (plotted in Figure 2)
observed at AIA 94 and 211 {\AA}, (top-panel).
The velocities are given by linear fitting to the distance-time profiles.}
\label{fig:fig4}
\end{figure}

\subsection{Combined Analysis}

\begin{figure*}[ht]
\includegraphics[width=0.95\linewidth]{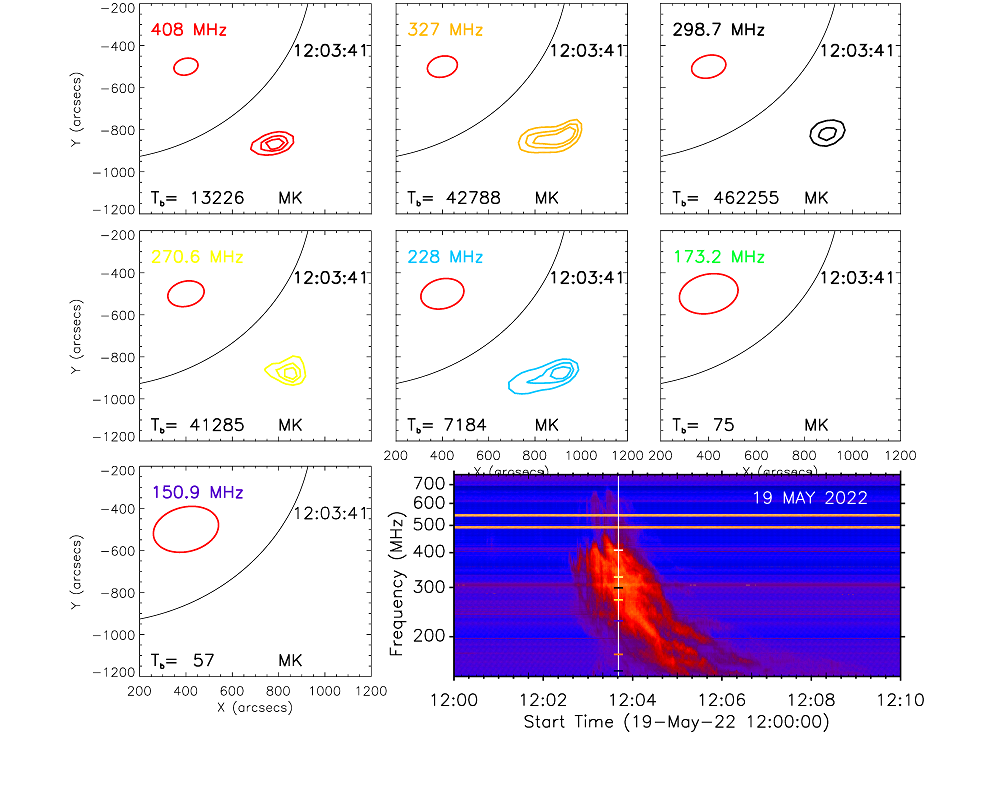}
\caption{Type$–$II radio sources observed by NRH at different
frequencies at 12:03:41 UT together with their dynamic spectrum.
The vertical white line represents the timing line. The radio
sources are represented by the 80 $\%$, 90 $\%$, and 95 $\%$
contours of the corresponding T$_{Bmax}$. The red ellipse represents the beam shape at different NRH frequencies. The beam size is about $\sim$ 53 arcsecs at 408 MHz, $\sim$ 66 arcsecs at 327 MHz, $\sim$ 73 arcsecs at 298.7 MHz, $\sim$ 80 arcsecs at 270.6 MHz, $\sim$ 95 arcsecs at 228 MHz, $\sim$ 125 arcsecs at 173.2 MHz and $\sim$ 144 arcsecs at 150.9 MHz along the major axis and is about $\sim$ 39 arcsecs at 408 MHz, $\sim$ 49 arcsecs at 327 MHz, $\sim$ 53 arcsecs at 298.7 MHz, $\sim$ 59 arcsecs at 270.6 MHz, $\sim$ 70 arcsecs at 228 MHz, $\sim$ 92 arcsecs at 173.2 MHz and $\sim$ 105 arcsecs at 150.9 MHz along the minor axis. The duration of the animation is 15s. (An animation of this figure is available).}
\label{fig:fig5}
\end{figure*}

In Figure 3, we have plotted the type-II sources
at 12:03:41 UT onto the EUV-WL bright images. The sources are associated with the bright
and dense loops underlying the streamer structure, as mentioned

In Figure 5 and the accompanying movie, we present
the radio sources observed by NRH with
the 80, 90, and 95$\%$ contour levels of the
corresponding maximum $T_B$ ($T_{Bmax}$, written on each panel) at
given NRH frequencies. Only sources with $T_{Bmax}$ $>$ 10$^8$ K
were presented to avoid interference from weaker emission.
The burst appears at 408 MHz after 12:02 UT; sources at
several frequencies appear together during most time
of the burst due to its wide band feature. See
Figure 5 for one such moment at 12:03:41 UT with the
type-II sources extending from 408 to 228 MHz.
The NRH sources present a gradual outward motion,
in line with the slowly-drifting feature of the spectra.

\begin{figure*}
\includegraphics[width=0.95\linewidth]{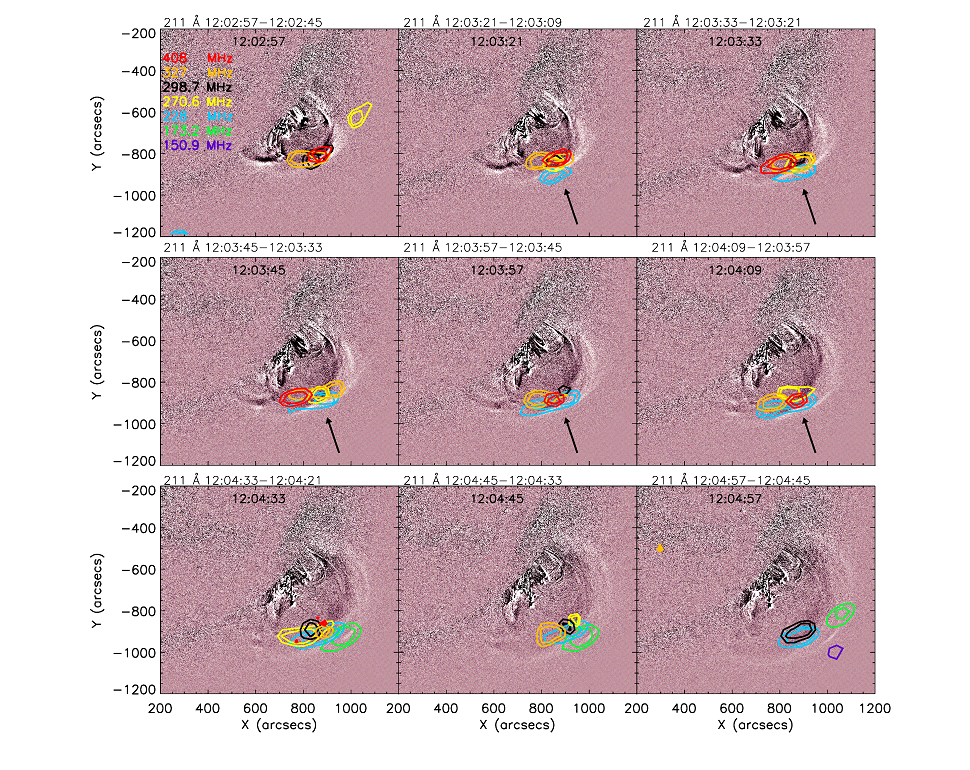}
\caption{The temporal evolution of NRH radio sources, superposed onto the closest-in-time AIA 211 {\AA}, images. The radio sources are represented by 90 $\%$, and 95 $\%$ of T$_{Bmax}$ contours. The duration of the animation is 29s. (An animation of this figure is available).}
\label{fig:fig6}
\end{figure*}

To further reveal the relation of NRH sources
with the EUV shock wave, in Figure 6 we overplot
the 90 and 95$\%$ contours of T$_{Bmax}$
onto the closest-in-time running difference images of AIA 211 {\AA}.
We made four observations: (1) the type-II sources are
co-spacial with the nose front of the EUV shock wave structure,
(being separated slightly by $\sim$ 0.01 $R_\odot$);
(2) the sources basically overlap with each other before 12:04 UT
(without spatial separation of $\sim$ 0.001 $R_\odot$);,
later they become spatially dispersed with lower-frequency sources
being further away from the disk; (3) as mentioned the sources
at several frequencies can appear simultaneously, they distribute
within a broad region of $\sim$ 150 $\times$ 200 arcsecs and
(4) during most time of the burst its sources are spatially dispered
yet centering around the dip region of the shock front according
to Figure 6; and the dip corresponds to the transit of the CME shock
across the bright dense loop tops according to Figures 3 and 6.

\section{Conclusions and Discussion}

Here we report an unusual high-frequency wide-band
type-II solar radio burst. It starts from $\sim$
600-700 MHz, with an instantaneous bandwidth of $\sim$ 300 MHz,
and the relative bandwidth is as large as $\sim$ 1 $\pm$ 0.05.
The burst sources distribute
within a broad region of $\sim$ 150 $\times$ 200 arcsecs, and
center around a dip region along the EUV shock wave front.

The dip is likely due to the strong compression caused by the
shock propagation into bright and dense loops
where the Alfv{\'e}nic speed is relatively low so is the
shock speed. According to the EUV and WL data, bright and dense loops that connect
the two spots are spacially correlated with
the shock dip, indicating the shock has transitted across
them with strong interaction/compresssion that leads to
the type-II burst. In this case,
the generation of efficient electron acceleration and type-II
bursts should be related to the shock transit across these magnetically-closed
loop structures. This is in line with a series of numerical studies by
Kong et al. (2015, 2016). To illustrate such
scenario, we present a sketch in Figure 7 to show
the shock transit across the loop system, the accompanying
shock and its dip, the type-II sources, as well
as the overlying large-scale helmet streamer.

The burst has a high starting frequency since it originates from
a dense region in the corona. This is consistent with the following
two aspects of our observations: (1) according to the AIA 94 and 211 {\AA} data,
the EUV shock-like structure get steepened at $\sim$ 1.23 R$_\odot$
that is very close to the solar surface; (2) the type-II source
coincides with the shock dip that forms since the shock transits
a region denser than surroundings (along the shock front), where the Alfv{\'e}nic
speed becomes smaller and so does the local propagation speed of the shock.

The wide-band feature of the type-II burst means that the
sources stem from a region with large density variation,
either the source contains large amplitude of density
fluctuation or the source extends over a broad region with
large inhomogeneity. Our observations favor the latter scenario
that the wide-band type-II burst stems from the broad sources centering
around the shock dip, meaning larger-than-usual plasma densities
been sampled by the coherent emission process.

Unfortunately, we cannot tell whether the burst belongs to the fundamental
or harmonic branch with the data available. Yet, according to it high
starting frequency, the strong scattering/absorption effect of the fundamental
plasma emission (Melrose, 1970),
also earlier studies indicating that in the corona the harmonic branch is
usually stronger (Roberts, 1959; Cliver et al. 1998; Gopalswamy, 2000; Chernov \& Fomichev, 2021 and Pohjalainen et al. 2021), we suggest that the burst represents the harmonic branch.
At 12:03:41 UT, the upper and lower frequencies of the burst are $\sim$ 440 and 180 MHz, respectivley.
Then the type-II source densities at 12:03:41 UT should vary from $\sim$ 10$^{9}$ to 10$^{8}$ cm$^{-3}$.
Note that the ratio of these densities is about 10,
much larger than the largest-possible compression ratio of
an magnetohydrodynamic shock. This indicates
the wide-band feature is not due to density variations
across the shock layer. Further studies on similar wide-band
type-II bursts are desired for deeper understanding of their origin.

\begin{figure}
\includegraphics[width=\hsize]{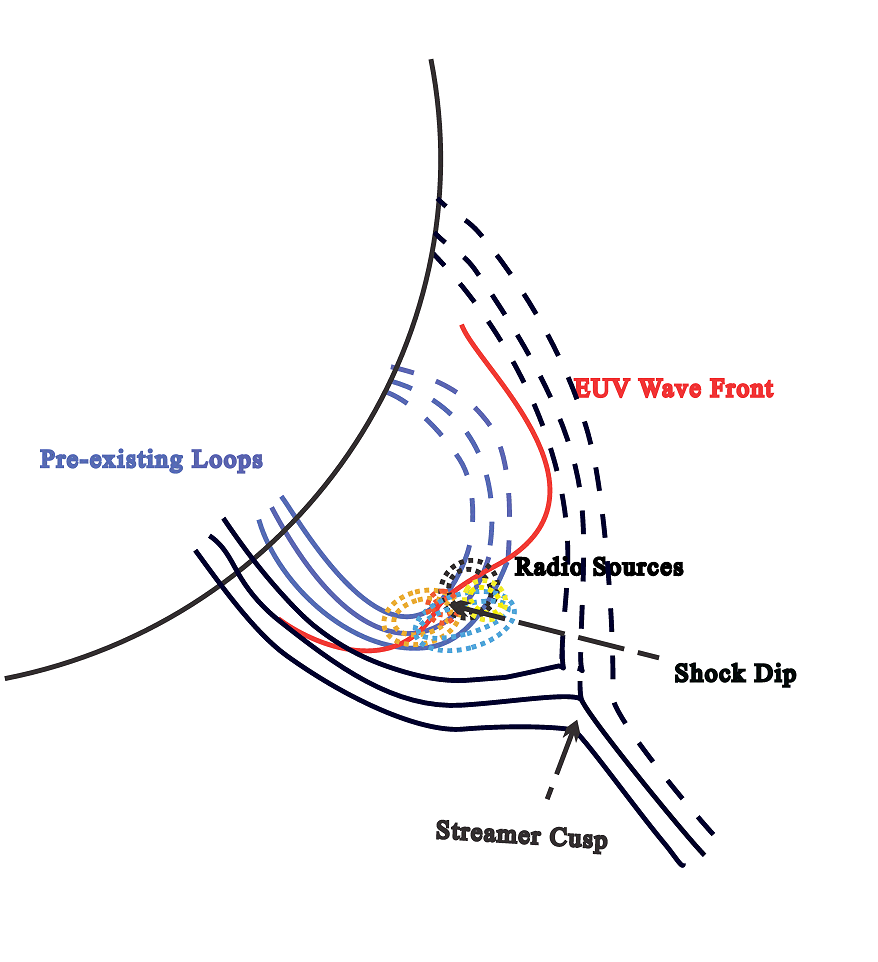}
\caption{The sketch of the shock-loop system with the dip region and
the corresponding type-II radio sources.}
\label{fig:fig7}
\end{figure}

\begin{acknowledgements}
      This work was supported by POB Anthropocene research program of Jagiellonian University, Krakow, Poland.
Yao Chen acknowledges the NNSFC grant 42127804. G. Michalek acknowledges the support under Grant 2023/49/B/ST9/00142 by the National Science Centre, Poland. We thank the ORFEES, NRH, SOHO, NOAA teams for providing
the data.
\end{acknowledgements}

%
%


\end{document}